\documentclass[preprint2]{emulateapj}
\usepackage{psfig}

\catcode`\@=11
\def\gsim{\ifmmode{\,\mathrel{\mathpalette\@versim>\,}}
    \else{$\,\mathrel{\mathpalette\@versim>}\,$}\fi}
\def\lsim{\ifmmode{\,\mathrel{\mathpalette\@versim<\,}}
    \else{$\,\mathrel{\mathpalette\@versim<}\,$}\fi}
\def\@versim#1#2{\lower 2.9truept \vbox{\baselineskip 0pt \lineskip
    0.5truept \ialign{$\m@th#1\hfil##\hfil$\crcr#2\crcr\sim\crcr}}}
\catcode`\@=12  
\def\km{{\rm \,km}}

\def\sminus{{\rm \,s^{-1}}}

\def\kpc{{\rm \,kpc}}

\def\fdmp{f_{\rm dm}^{\rm p}}

\def\Conc{C_{200}}

\def\Mstar{M_*}
\def\Mstartilde{\tilde{M}_*}
\def\Mstarfin{M_{\rm *,fin}}
\def\Mdmfin{M_{\rm dm,fin}}
\def\Nstar{N_*}
\def\Ntot{N_{\rm tot}}
\def\Ndm{N_{\rm dm}}
\def\vstartilde{\tilde{v}_*}
\def\tstartilde{\tilde{t}_*}

\def\dz{d_0}

\def\Msun{M_{\odot}}

\def\Metp{M_{\rm e2}^{\rm p}}
\def\Mtot{M_{\rm tot}}

\def\Mdm{M_{\rm dm}}

\def\Mone{M_1}
\def\Mtwo{M_2}

\def\Psistar{\Psi_*}
\def\Psidm{\Psi_{\rm dm}}

\def\rhostar{\rho_*}

\def\rhotot{\rho_{\rm tot}}
\def\rhototzero{\rho_{\rm tot,0}}
\def\rhodm{\rho_{\rm dm}}

\def\Mdmzero{{M}_{\rm dm,0}}
\def\dzero{d_{0}}
\def\vzero{v_{0}}
\def\vzeropar{v_0^{\parallel}}
\def\vzeroperp{v_0^{\perp}}
\def\dvzero{{\bf d}_{0}}
\def\vvzero{{\bf v}_{0}}

\def\Lv{{\bf L}}

\def\Re{R_{\rm e}}

\def\Dt{\Delta t}

\def\sae{a_{\rm e}}
\def\sbe{b_{\rm e}}

\def\rs{r_{\rm s}}
\def\rt{r_{\rm t}}
\def\rvir{r_{\rm vir}}

\def\rstartilde{\tilde{r}_{*}}
\def\fstar{f_*}

\def\ra{r_{\rm a}}
\def\rM{r_{\rm M}}
\def\rMtot{r_{\rm M,tot}}
\def\am{a_{\rm m}}
\def\bm{b_{\rm m}}

\def\sigmav{\sigma_{\rm v}}

\def\sget{\sigma_{\rm e2}}
\def\sgetsq{\sigma_{\rm e2}^2}

\def\cet{c_{\rm e2}}
\def\Sget{\Sigma_{\rm e2}}

\def\gammap{\gamma^{\prime}}

\def\en{{\mathcal{E}}}
\def\Psitot{\Psi_{\rm tot}}
\def\hatE{\hat{E}}
\def\hatL{\hat{L}}

\def\alphar{\alpha_{R}}
\def\alphac{\alpha_{c}}
\def\alphas{\alpha_{\sigma}}

\def\const{\mathit{const}}
\def\thetamin{\theta_{\rm min}}
\def\rhomax{\rho_{\rm max}}

\slugcomment{Accepted, August 12, 2009}

\shorttitle{Dry mergers of early-type galaxies}
\shortauthors{Nipoti et al.}

\begin{document}


\title{Dry mergers and the formation of early-type galaxies:
  constraints from lensing and dynamics}


\author{C.~Nipoti}
\affil{Dipartimento di Astronomia,
Universit\`a di Bologna, via Ranzani 1, 40127 Bologna, Italy}
\email{carlo.nipoti@unibo.it}
\author{T.~Treu}
\affil{Department of Physics, University of California, Santa Barbara, CA
93106-9530, USA}
\and
\author{A.~S.~Bolton}
\affil{Institute for Astronomy, University of Hawaii, 2680 Woodlawn Dr., Honolulu, HI 96822 USA}




\begin{abstract}
Dissipationless (gas-free or ``dry'') mergers have been suggested to
play a major role in the formation and evolution of early-type
galaxies, particularly in growing their mass and size without altering
their stellar populations.  We perform a new test of the dry merger
hypothesis by comparing N-body simulations of realistic systems to
empirical constraints provided by recent studies of lens early-type
galaxies.  We find that major and minor dry mergers: i) preserve the
nearly isothermal structure ($\rho_{\rm tot}\propto r ^{-2}$) of
early-type galaxies within the observed scatter; ii) do not change
more than the observed scatter the ratio between total mass $M$ and
``virial'' mass $\Re\sgetsq/2G$ (where $\Re$ is the half-light radius
and $\sget$ the projected velocity dispersion); iii) increase
  strongly galaxy sizes ($\Re\propto M^{0.85\pm0.17}$) and weakly
  velocity dispersions ($\sget\propto M^{0.06\pm0.08}$) with mass,
thus moving galaxies away from the local observed $M$-$\Re$ and
$M$-$\sget$ relations; iv) introduce substantial scatter in the
$M$-$\Re$ and $M$-$\sget$ relations.  Our findings imply that
  ---unless there is a high degree of fine tuning of the mix of
  progenitors and types of interactions--- present-day massive
  early-type galaxies cannot have assembled more than $\sim 50\%$ of
  their mass, and increased their size by more than a factor $\sim
  1.8$, via dry merging.
\end{abstract}

%




\section{Introduction}

Although early-type galaxies appear to be relatively simple in terms
of their internal structure and stellar populations, their formation
and evolution are still poorly understood.  In the hierarchical model
of structure formation, galaxy mergers play a central role in the
formation and evolution of early-type galaxies. Many pieces of
evidence are consistent with the merging hypothesis \citep[see
  e.g.,][and references therein]{vanDokkum05}. However, other
observational facts appear hard to reconcile with substantial amounts
of merging. Among the facts that have been traditionally used to argue
against the central role of mergers are the high observed central
density of early-type galaxies \citep{Ostriker80,Carlberg86}; the
small scatter in empirical scaling laws such as the color magnitude
relation \citep{Bower92}, the Fundamental Plane \citep[hereafter
  FP;][]{Dre87,Djo87}, and the relation between black hole mass
$M_{\rm BH}$ and host galaxy mass or central stellar velocity
$\sigma_0$ dispersion \citep{Ferrarese2000,Geb00,Marconi03,Haring04};
and the correlation between abundance ratios and host galaxy
properties \citep[e.g.,][]{P+M08}.

Theoretical and observational studies
\citep[e.g.,][]{Ben92,Ciotti01,Robertson06,Ciotti07,Scarlata07,Faber07}
have argued that a combination of dissipative and dissipationless
processes is needed to match the observed properties of early-type
galaxies. Dissipation is needed for example to create the high density
regions and close to isothermal density profiles, while
dissipationless (or ``dry'') mergers are needed to grow the most
massive galaxies while preserving the uniformly old stellar
populations and the structural properties of present-day galaxies
\citep[e.g.][]{Naab07}. In particular, both theories and observations
have suggested that a large fraction of the immediate progenitors of
today's massive red galaxies could be spheroidal galaxies themselves
\citep[e.g.,][]{vanDokkum99, Kho03,Naab06,Bell06,Faber07}. Thus, even
though mergers may not be the only process at work
\citep[e.g.,][]{Bundy07}, dry mergers might play an important role,
especially at relatively recent times, i.e. below redshift $\sim1.5$,
after the bulk of star formation is completed for massive early-type
galaxies in the local Universe \citep[e.g.,][]{Trager00,
  Tre05,Renzini06}.

In this paper, we take advantage of a new set of constraints provided
by strong gravitational lensing studies to test the viability of the
dry-merger scenario via a set of numerical simulations.  In practice,
we first construct realistic models of early-type galaxies so that
they obey known scaling relations between their structural and
kinematic properties, and then study how these properties change as a
result of dissipationless mergers.

Our approach is similar to that followed by \citet{Nip03a},
\citet{Gonz03} and \citet[][see also Kazantzidis et al.\ 2005,
    Robertson et al.\ 2006 and Johansson, Naab, \& Burkert, 2009a for
    studies that include dissipation]{Boy06}, who tested how the FP,
its projections and the $M_{\rm BH}$-$\sigma_0$ relation are affected
by mergers. The bottom line of these studies is that dry merging
between progenitor spheroidal galaxies that lie on the FP,
\citet{Kor77} and \citet{Fab76} relations produce spheroidal galaxies
that are close to the edge-on FP, but deviate from the Kormendy and
Faber-Jackson relations, because they have too large effective radius
and too low central velocity dispersion.

Taking a step further, we focus here on lensing-derived properties,
testing the results of our simulations against: i) isothermality of
the total mass density profile; ii) the correlations between lensing
mass and size or stellar velocity dispersion; iii) the Mass Plane,
i.e. the correlation between surface mass density, effective radius
and velocity dispersion discovered by strong lensing studies.  As
discussed by \citet{Bol07,Bol08b}, the lensing constraints add a
qualitatively new piece of information---an accurate aperture mass for
each galaxy---that helps circumvent some of the problems inherent to
previous studies.  The comparison of N-body simulation with
traditional scaling relations is not straightforward, because it
relies on assumptions on the mass-to-light ratio to convert luminosity,
which enters the observed correlations, into mass. Moreover, among the
quantities involved in the traditional empirical correlations (e.g.,
galaxy luminosity, size, and velocity dispersion), only velocity
dispersion is affected by the galaxy dark matter (hereafter DM)
content and distribution.  Information coming from gravitational
lensing, and thus constraining the total galaxy mass distribution,
effectively solves this problem.

The paper is organized as follows. First, in \S~\ref{sec:obs} we
review the lensing constraints on the structure of early-type
galaxies. Then, in \S~\ref{secsim}, we describe the seed-galaxy
models, initial orbital configurations, and numerical methods that we
use to simulate the dry-merging process.  In \S~\ref{sec:tden} and
\S~\ref{sec:lens} we present the results of comparisons between our
merger simulations and observations in terms of the effect of dry
mergers on density profiles, central DM fractions and scaling
relations of galaxies throughout the merging hierarchies.  In
\S~\ref{sec:highz} we discuss our results in the context of the
redshift evolution of early-type galaxies.  Finally, \S~\ref{seccon}
presents our conclusions.

\begin{figure}
\epsscale{1.0}
\plotone{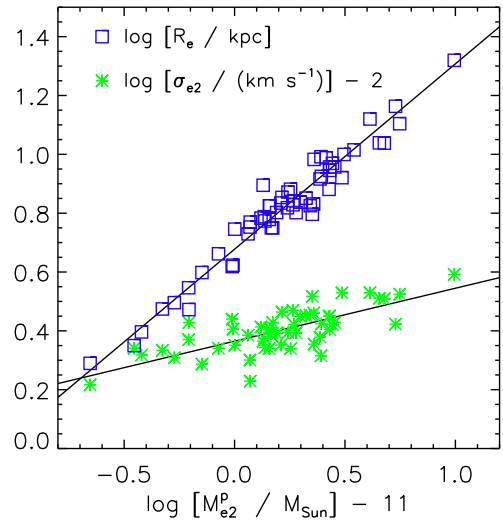}
\caption{Mass-space analogs of the Kormendy and Faber-Jackson
relations as defined by data on the SLACS early-type gravitational lens
sample from \citet{Bol08a}.  Squares show the variation of
half-light radius $\Re$ with lensing mass $\Metp$; stars show the
variation of velocity dispersion $\sget$ with $\Metp$.
\label{rvm}}
\end{figure}

\begin{figure}
\epsscale{1.0}
\plotone{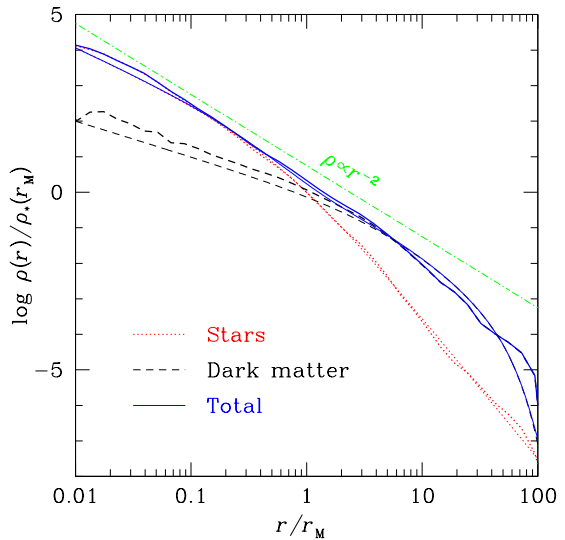}
\caption{Stellar, dark-matter and total density distributions of the
  seed galaxy (model A; thin curves) and of the end-product (thick
  curves) of the major-merger simulation 2A5po. $\rM$ is the
  angle-averaged half-mass radius of the stellar distribution.}
\label{fig:den}
\end{figure}

\begin{figure}
\epsscale{1.0}
\plotone{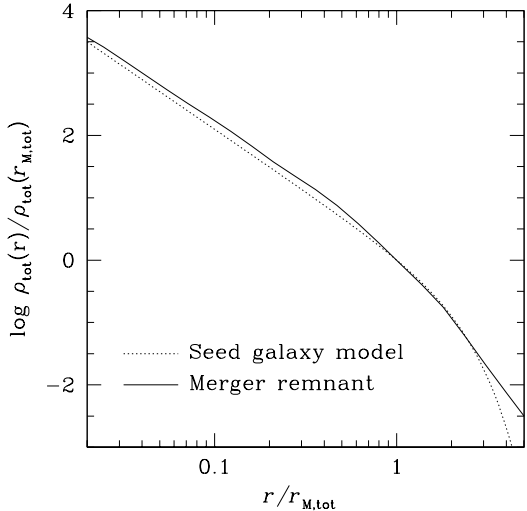}
\caption{Total density profiles of the seed galaxy model (dotted
  curve) and end-product (solid curve) of a parabolic head-on merging
  between two identical systems with truncated singular isothermal
  sphere total density distribution. $\rMtot$ is the angle-averaged
  half-mass radius of the total mass distribution.}
\label{fig:sis}
\end{figure}

\section{Lensing constraints on the structure of early-type galaxies}
\label{sec:obs}

The properties of massive early-type galaxies are strongly constrained
by the combination of kinematic and strong gravitational lensing
observations. In particular, there is growing evidence that these
systems have total mass density profile close to isothermal
$\rho\propto r^{-2}$ over a large radial interval \citep{Gav07}, with
an intrinsic scatter in logarithmic slope of less than $\sim 10$\%
\citep{Koo06,Koo09}.

In addition, \cite{Bol07,Bol08b}, using a sample of strong
gravitational lenses, have shown that early-type galaxies lie on a
Mass Plane (MP)
\begin{equation}
\log\Re=\am\log\sget+\bm\log\Sget+\const,
\label{eq:MP}
\end{equation}
where $\sget$ is the projected velocity dispersion within an aperture
radius $\Re/2$ and $\Sget$ is the surface mass density within $\Re/2$,
with $\am=1.82 \pm 0.19$, $\bm=-1.20 \pm 0.12$ and RMS orthogonal
scatter of 1.24 when normalized by the observational errors.
The MP is the gravitational lensing analogue of the
traditional FP.

The fact
that $(\am,\bm)$ are consistent with $(2,-1)$ and that the scatter is small
can be expressed in terms of structural and dynamical homology of the
lenses, by defining the dimensionless structure parameter
\begin{equation}
\cet\equiv{2G\Metp\over \Re\sgetsq},
\label{eq:cet}
\end{equation}
where $\Metp$ is the total projected mass within $\Re/2$.  For their
sample of lens early-type galaxies from the Sloan Lens ACS (SLACS)
Survey, \cite{Bol08b} find on average
\begin{equation}
\langle\log \cet\rangle=0.53\pm0.057.
\label{eqobsrange}
\end{equation}
We note that the observed scatter on $\langle\log \cet\rangle$ is
0.08, but here we consider the estimated intrinsic scatter 0.057
\citep[see][]{Bol08b}.  Importantly, \citet{Bol08b} find that $\cet$
has no systematic variation with galaxy mass: this is the sense in
which the lens galaxies form a homologous family.

In the present paper we also consider the analogues of the traditional
luminosity--size \citep{Kor77} and luminosity--velocity-dispersion
\citep{Fab76} relations, by deriving the projected-mass--size and the
projected-mass--velocity-dispersion relation for the SLACS sample of
lenses.

To derive total-mass versions of the Kormendy and Faber-Jackson
relations, we use the velocity-dispersion, half-light radius, and
lensing-mass measurements for the SLACS strong-lens sample as
published in \citet{Bol08a}.  We restrict our attention to the 57 lens
systems wherein the lens galaxy has early-type morphology and single
multiplicity.  For the mass--velocity dispersion relation, we further
restrict ourselves to the 53 systems for which the Sloan Digital Sky
Survey (SDSS) spectroscopy is of sufficient signal-to-noise to permit
a velocity-dispersion measurement.  \citet{Bol08a} provide lensing
measurements using both singular isothermal ellipsoid (SIE) and
light-traces-mass (LTM) models for the mass density of the foreground
lensing galaxies, which can be used to gauge the dependence of results
upon the radial mass-density profile used to extract the
strong-lensing aperture masses.

Taking the SIE aperture masses as our reference values, and fitting
for the relation that minimizes the orthogonal scatter in logarithmic
space, we find a mass--size relationship given by
\begin{eqnarray}
\nonumber &\log& \left({\Re \over \kpc}\right) = (0.63 \pm 0.02) \\
&\times& \log \left({\Metp\over 10^{11}\Msun}\right) + 0.68 \pm 0.01.
\label{eq:reff}
\end{eqnarray}
The observed scatter in $\log \Re$ at fixed mass is 0.051, which is
reduced to 0.048 by subtracting in quadrature
the error in $\log \Re$ as estimated by \citet{Bol08a} to
obtain an estimate of the intrinsic scatter in this quantity.
We note that there will be a degree of covariance between the
axes of this relation, since the
measured $\Re$ is used to evaluate the aperture-mass of
the SIE model at $\Re / 2$ to obtain $\Metp$.
Fitting similarly for the mass--velocity-dispersion relation, we find
\begin{eqnarray}
\nonumber &\log& \left({\sget \over 100\km\sminus}\right) = (0.18 \pm 0.02) \\
&\times& \log \left({\Metp\over 10^{11}\Msun}\right) + 0.36 \pm 0.01.
\label{eq:sigma}
\end{eqnarray}
This has an observed scatter in $\log \sget$ at fixed mass of 0.045,
implying an intrinsic scatter of 0.034.  Errors in all coefficients
are computed from the RMS scatter in fits to 1000 bootstrap-resampling
draws of 57 or 53 galaxies apiece.  Using LTM lensing aperture masses
instead of those from SIE models, the coefficients of both relations
change only insignificantly.  These relations are shown in
Fig.~\ref{rvm}.

As discussed in several papers \citep{Bol06,Tre06,Bol08a,Tre09} the
SLACS lenses are found to be indistinguishable from control samples of
SDSS galaxies with the same stellar velocity dispersion and size, in
terms of luminosity/surface brightness, location on the FP, and
distribution of environments.  We will thus assume that, within the
range of observational errors, results found for the lens sample,
including the MP, are generic properties of the overall class of
massive early-type galaxies.

\section{Dissipationless merging simulations}
\label{secsim}

We ran N-body simulations of dissipationless galaxy merging, exploring
both major and minor mergers. In the case of major mergers we simulate
two colliding systems, which are equal-mass, identical galaxy models
with stellar and DM components, and no gas.  In major-merger
  hierarchies, galaxies grow in mass by subsequent binary equal-mass
  mergers, up to an increase in stellar mass of a factor of $\sim 8$.
In the first step of a hierarchy the seed galaxies are spherically
symmetric and non-rotating.  In the second step of a hierarchy the
end-product of the first step is replicated, and two such galaxies
encounter one another.  At this stage, the systems are no longer
spherically symmetric, thus each of them is initially randomly rotated
around its center of mass. The setup of the initial conditions of the
third step are obtained in the same way, starting from the remnant of
the second step.

In the case of minor mergers, we simulate the accretion of several
smaller satellites onto a more massive system (the mass ratio between
each satellite and the massive galaxy is either $1/10$ or $1/5$).  For
a galaxy to grow significantly on reasonably short timescales through
minor mergers, more than one satellite is expected to interact with
the galaxy at the same time. For this reason, we considered
hierarchies of minor mergers in which each step is represented by a
multiple merger, so the initial conditions represent a massive galaxy
surrounded by five or ten infalling satellites. Each galaxy collision
is followed up to the virialization of the resulting stellar system. A
fraction of the particles is unbound at the end of the simulation: we
define the merging end-product as the system composed by the bound
stellar and DM particles.  We note that in the case of minor-merger
hierarchies not all the satellites present in the initial conditions
must necessarily merge with the central galaxy.  In some cases one or
two of the satellites are found on an unbound orbit when all the other
systems have already merged and the merger remnant relaxed. When this
happens, the satellites are considered escaped and the merging is thus
characterized by a significant stellar and DM mass loss.

Altogether we simulated 12 merging hierarchies: 6 major-merger
hierarchies of 3 steps each, 2 major-merger hierarchies of 2 steps
each, and 4 minor-merger hierarchies of 2 steps each.

\subsection{Seed galaxy models}
\label{secmod}

The seed galaxy models are spherically symmetric equilibrium
two-component systems, with stellar and DM components. The stellar
density distribution of the seed galaxies is represented by a $\gamma$
model \citep{Deh93,Tremaine94}:
\begin{equation}
\rhostar (r)= {3-\gamma\over 4\pi}{\Mstartilde\rstartilde \over r^{\gamma} (\rstartilde+r)^{4-\gamma}}  
\qquad  (0\leq \gamma <3),
\end{equation}
%
%
where $\Mstartilde$ is the total stellar mass. We assume $\gamma=1.5$,
thus our models are centrally steeper than \citet{Her90} models
($\gamma=1$) and shallower than \citet{Jaf83} models ($\gamma=2$).  We
use $\rstartilde$ and $\Mstartilde$ to define time and velocity units
$\tstartilde=(\rstartilde^3/G\Mstartilde)^{1/2} $ and $\vstartilde =
\rstartilde/\tstartilde = (G\Mstartilde/\rstartilde)^{1/2}$.  The
DM halo is described by a \citet[][NFW]{Nav96} model, so the
DM density distribution is
\begin{equation}
\rhodm (r)={\Mdmzero \over r(r+\rs)^2}\exp\left[{-\left({r\over \rvir}\right)^2}\right],
\label{eqrhodm}
\end{equation}
where $\rs$ is the scale radius, $\Mdmzero$ is a reference mass and we
adopt an exponential cut-off to truncate the distribution smoothly at
the virial radius $\rvir$, so the  total DM mass 
$\Mdm=4\pi \int_0^{\infty}\rhodm(r)r^2 dr$
is finite.  We assume Osipkov-Merritt \citep[][]{Osi79,Mer85}
anisotropy in the velocity distribution of the stellar component,
whose distribution function is then given by
\begin{equation}
f(Q)=\frac{1}{\sqrt{8}\pi^2}\frac{d}{dQ}
       \int_0^Q{\frac{d{\varrho_*}}{d\Psitot}}{\frac{d\Psitot}{\sqrt{Q-\Psitot}}},
\label{eqdf}
\end{equation}
where
\begin{equation}
\varrho_* (r)=\left(1+\frac{r^2}{\ra^2}\right)\rhostar (r).
\end{equation}
The variable $Q$ is defined as $Q\equiv \en-{L^2/2\ra^2}$, where the
relative (positive) energy is given by $\en =\Psitot-v^2/2$, $v$ is
the modulus of the velocity vector, the relative (positive) total
potential is $\Psitot=\Psistar +\Psidm$ ($\Psistar$ and $\Psidm$ are,
respectively, the relative potentials of the stellar and DM
components), $L$ is the angular momentum modulus per unit mass, and
$f(Q)=0$ for $Q\leq0$. The quantity $\ra$ is the so--called anisotropy
radius: for $r \gg\ra$ the velocity dispersion tensor is mainly
radially anisotropic, while for $r \ll \ra $ the tensor is nearly
isotropic.  In the limit $\ra \to\infty$, $Q=\en$ and the velocity
dispersion tensor becomes globally isotropic.

The orbital distribution of the DM halo is assumed isotropic, so the
distribution function of the DM component is given by
equation~(\ref{eqdf}) where $Q=\en$ and $\rhodm(r)$ substitutes
$\varrho_*(r)$.  The total (stars plus DM) density profile is
$\rhotot (r)=\rhodm(r)+ \rhostar(r)$
and $\Mtot(r)=4\pi \int_0^{r}\rhotot(r')r'^2dr'$ is the total mass
within $r$.  This family of $\gamma=1.5$ plus NFW models have four
free parameters: concentration $\Conc\equiv \rvir/\rs$, the stellar
mass fraction $\fstar\equiv\Mstartilde/(\Mstartilde+\Mdm)$, the ratio
$\xi\equiv\rs/\Re$ and the anisotropy radius $\ra$.  We consider
  different choices of these parameters for the seed galaxies of our
  hierarchies.  In what we refer to as model A (see
  Table~\ref{tab:mod}), we assume $\Conc=7$, $\fstar=0.09$, $\xi=7.8$,
  so that the resulting total mass profile is close to isothermal
  ($\rho_{\rm tot}\propto r ^{-2}$) in the radial range $0.05 \lsim
  r/\rM \lsim 10$ (see thin solid curve in Fig.~\ref{fig:den}), where
  $\rM$ is the radius of the sphere containing half of the stellar
  mass, and $\ra/\rstartilde=1.4$, so it lies exactly on the MP ($\log
  \cet \simeq 0.53$).  We also consider models with $\Conc=7$,
  $\fstar=0.09$, $\xi=7.8$, but different values of $\ra/\rstartilde$:
  an isotropic model ($\ra/\rstartilde=\infty$; model B) and a more
  radially anisotropic model ($\ra/\rstartilde=0.7$; model C).  In
  other hierarchies the seed galaxy is more DM dominated (model D): a
  $\gamma=1.5$ plus NFW model with $\Conc=8$, $\fstar=0.02$,
  $\xi=11.6$ and $\ra/\rstartilde=1$ so that also this model has
  close-to-isothermal total density profile and has $\log \cet \simeq
  0.53$.

Our seed galaxy models are primarily constructed to reproduce the
total density distribution and structural parameter $\cet$ of massive
($\Mstar\sim 10^{11}\Msun$) local early-type galaxies, as constrained
by strong gravitational lensing and kinematics; we did not attempt a
full exploration of the space of parameters $\Conc$, $\fstar$ and
$\xi$, but we adopted values that are expected to bracket the ranges
expected on the basis of theoretical investigations \citep[see][and
  references therein]{Nip08}. Additional observational constraints on
the DM content of present-day early-type galaxies come from weak
gravitational lensing surveys \citep{Man06,Gav07}. According to the
weak-lensing results of \cite{Man06} the stellar mass fraction
$\fstar$ of $\Mstar\sim10^{11}\Msun$ early-type galaxies are of the
order of $0.03-0.06$ (if uncertainties on the Initial Mass Function
are taken into account), so the values of $\fstar$ of our seed galaxy
models bracket the observationally estimated range of $\fstar$ for
systems of this mass. However, $\fstar$ estimated from weak-lensing
tend to decrease with stellar mass above $\Mstar\sim10^{11}\Msun$
\citep{Man06}, so the more DM dominated model D ($\fstar=0.02$) should
be considered more realistic than model A to represent more massive
systems.

To address the question of whether $r^{-2}$ density profiles are
preserved by merging we also ran an equal-mass merger simulation in
which each seed galaxy is a one-component system with total density
profile
\begin{equation}
\rhotot (r)= \rhototzero \left({\rt \over r}\right)^2 \exp\left[{-\left({r\over \rt}\right)^2}\right],
\label{eqrhosis}
\end{equation}
where $\rhototzero$ is a reference density. The density distribution
in the equation above is that of a singular isothermal sphere (SIS),
smoothly truncated at a radius $\rt$ so that the total mass
$\Mtot=4\pi \int_0^{\infty}\rhotot(r)r^2 dr$
is finite.  This truncated SIS model is assumed isotropic.

\subsection{Major-merging simulations}
\label{sec:maj}

In the case of major-merging hierarchies, in each step we follow the
evolution of a binary galaxy encounter, characterized by the
properties of the colliding systems and of their mutual orbit, which
we parametrize with the standard two-body approximation. For an
encounter between two galaxies of total masses $\Mone$ and $\Mtwo$ we
define the orbital energy per unit mass
\begin{equation}
E={1\over 2}\vzero^2-{G (\Mone+\Mtwo)\over \dzero},
\end{equation}
and the orbital angular momentum per unit mass
$\Lv=\dvzero\times\vvzero$,
where $\dvzero$ and $\vvzero$ are the relative separation and velocity
vectors of the centers of mass of the two colliding systems at the
initial time of the simulation. We ran simulations of parabolic
($E=0$), elliptic ($E<0$) and hyperbolic ($E>0$) orbits, and we
explored both head-on encounters ($L=0$) and off-center encounters
($L>0$).  In the case of encounters between two equal-mass, identical
galaxies it is useful to define the dimensionless orbital energy
$\hatE\equiv 2E/\sigmav^2$ and angular momentum modulus $\hatL\equiv
L/\sigmav\rMtot$, where $\sigmav$ is the virial velocity dispersion
and $\rMtot$ is the angle-averaged half-mass radius of the total
density distribution of the isolated systems~\citep{Bin87}.  We
consider six different major merging hierarchies, with seed galaxy
model A, in which the orbital parameters $\hatE$ and $\hatL$ are the
same in each step of the hierarchy. Only in the last step (run 8A1ho)
of the off-center hyperbolic major-merger hierarchy $\hatE$ and
$\hatL$ are smaller than in the previous steps, in order to avoid a
too long merging timescale.  In addition we explore two parabolic
hierarchies having as seed galaxy model D (one with $L=0$, the other
with $L>0$). The parameters of the simulations are summarized in
Tables~\ref{tab:mod} and \ref{tab:sim}.

\subsection{Minor-merging simulations}
\label{sec:min}

We present four minor-merging hierarchies, each consisting of two
steps of (multiple) minor-merging simulations.  Let us focus first on
the hierarchy represented by the two simulations 2A1m10x and 4A1m10x:
in the first step of this hierarchy the initial conditions consist of
a central galaxy model A, of stellar mass $\Mstartilde$ and scale
radius $\rstartilde$, surrounded by ten identical satellites, also
represented by model A, but with stellar mass $\Mstartilde/10$ and
scale radius $0.23 \rstartilde$.  The satellites are randomly
distributed around the central galaxy with relative distance between
each satellite and the central galaxy in the range
$100-140\,\rstartilde$ and with velocity directed towards the center
of mass of the central galaxy. The relative speed between each
satellite and the central galaxy is such that the orbit would be
parabolic in the absence of the other satellites.  The end-product of
run 2A1m10x is used as central galaxy of the initial conditions in the
second step of the hierarchy, in which it is surrounded by ten
identical satellites of mass $\Mstartilde/5$ and scale radius
$0.36\rstartilde$ with distribution analogous to run 2A1m10x, but now
with relative distance with respect to the central galaxy in the range
$200-280\,\rstartilde$.  Though each satellite would be on a radial
parabolic orbit in the absence of the other satellites, the latter act
as perturbers, so the orbits are actually neither strictly radial nor
strictly parabolic.

The set-up of the other minor-merging hierarchies are similar to the
one described above.  The hierarchy (2A1m10y, 4A1m10y) is just another
realization of the same set-up: the initial conditions differ just in
the relative distribution in phase space of the centers of mass of the
satellites. In the hierarchy (2A1m5, 4A1m5) we adopt five satellites
per step instead of ten, with stellar mass $0.2 \Mstartilde$ ($0.4
\Mstartilde$) and scale radius $0.36 \rstartilde$ ($0.56 \rstartilde$)
in the first (second) step.  In the hierarchy (2D1m10, 4D1m10) the
initial central galaxy and the ten satellites are represented by model
D, with same scale radii and masses as (2A1m10x, 4A1m10x), but
relative distance between each satellite and the central galaxy in the
range $170-230\,\rstartilde$ (first step) and $340-460\,\rstartilde$
(second step), on account of the more extended DM distributions.

\subsection{Numerical methods}
\label{secnum}
 
For the simulations we used the parallel N-body code FVFPS
\citep[Fortran Version of a Fast Poisson Solver;][]{Lon03,Nip03a},
based on \citet{Deh02} scheme. We adopt the following values for the
code parameters: minimum value of the opening parameter
$\thetamin=0.5$ and softening parameter $\varepsilon=0.03-0.1$ in
units of $\rstartilde$ (depending on the number of particles and on
the size of the merging systems). The time-step $\Dt$, which is the
same for all particles, is allowed to vary adaptively in time as a
function of the maximum particle density $\rhomax$: in particular, we
adopted $\Dt=0.3/(4\pi G \rhomax)^{1/2}$.

Numerically, the initial distribution of the seed galaxy particles in
phase space is realized as described in \citet{Nip02}.  While the
gravitational potential $\Psistar$ is analytic, $\Psidm$ is computed
by numerical integration, given the spherically symmetric density
distribution~(\ref{eqrhodm}).  We verified that the seed galaxy models
are in equilibrium by evolving them in isolation for several dynamical
times.  As a rule, in the first step of each merging hierarchy we use
$\Nstar \simeq 4 \times 10^4$ stellar particles and $\Ndm\simeq 4
\times 10^5$ (when $\fstar=0.09$) or $\Ndm\simeq 2 \times 10^6$ (when
$\fstar=0.02$; halo and stellar particles have the same
mass). Clearly, the number of particles increases with the hierarchy
steps, up to $\Ntot=\Ndm+\Nstar\simeq1.7-3.9\times10^6$ in the last
steps considered.

The intrinsic and projected properties of the merger remnants are
determined following \citet{Nip02,Nip06}.  In particular, we measure
the axis ratios $c/a$ and $b/a$ of the inertia ellipsoid (where $a$,
$b$ and $c$ are the major, intermediate and minor axis), and the
angle-averaged density distribution of the stellar and DM
components. We also measure the angle-averaged half-mass radius of the
stellar density distribution ($\rM$) and of the total density
distribution ($\rMtot$). For each end-product, in order to estimate
the importance of projection effects, we consider 50 random
projections and we use them to compute average projected quantities
and associated $1$-$\sigma$ scatter.  For each projection we measure
the ellipticity $\epsilon=1-\sbe/\sae$, the circularized projected
density profile and the circularized effective radius
{$\Re\equiv\sqrt{\sae\sbe}$} (where {$\sae$} and {$\sbe$} are the
major and minor semi-axis of the effective isodensity ellipse).  In
addition, we measure the central projected velocity dispersion
$\sget$, obtained by averaging the projected velocity dispersion over
the circularized surface density profile within an aperture of
$\Re/2$. The projected mass $\Metp$ is computed by counting stellar
and DM particles within the same projected elliptical aperture with
circularized radius $\Re/2$.

As a convergence test, we reran three of the simulations at higher
resolution (with five times more particles), finding good agreement
with the lower-resolution cases. In particular, the runs 2A5ph, 2A5po
and 2A5m10x differ from the runs 2A1ph, 2A1po and 2A1m10x,
respectively, just for the number of particles of the seed galaxies
(see Tables~\ref{tab:mod} and \ref{tab:sim}): the quantities $\Re$,
$\sget$ and $\Metp$, which are crucial for the present study, are
indistinguishable in lower and higher resolution simulations.

\begin{figure*}
\epsscale{1.}
\plotone{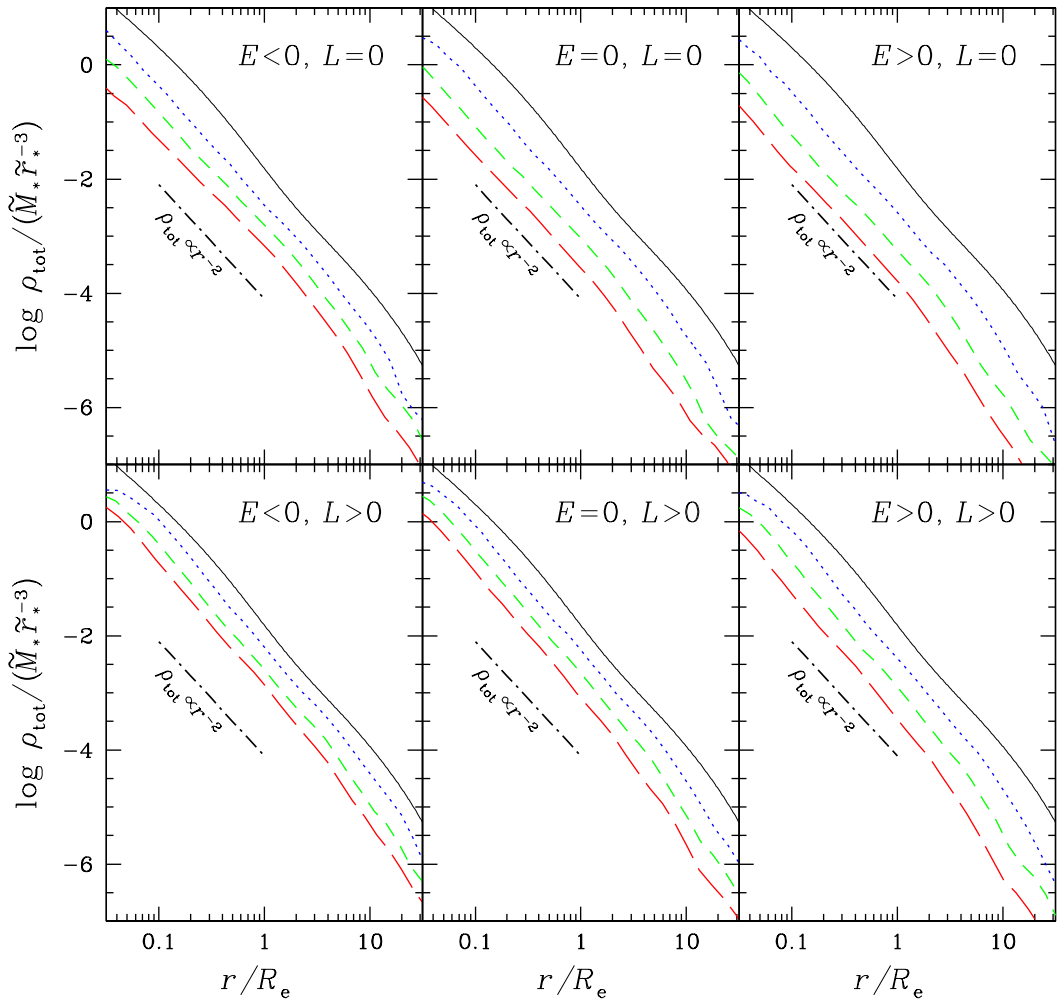}
\caption{Total density distributions of the seed galaxy (solid curves)
  and of the end-products of six major-merging hierarchies having the
  same seed galaxy (model A), but different orbital parameters. $\Re$
  is the average of the effective radius over 50 random lines of
  sight. The end-products of the first, second and third steps of the
  hierarchies are represented by dotted, short-dashed and long-dashed
  curves, respectively.  The dash-dotted line represents the
  isothermal distribution ($\rho_{\rm tot}\propto r^{-2}$) in the
  radial range $0.1\leq r/\Re \leq 1$, over which we fit the
  profiles.}
\label{fig:tden}
\end{figure*}

\begin{figure}
\epsscale{1.0}
\plotone{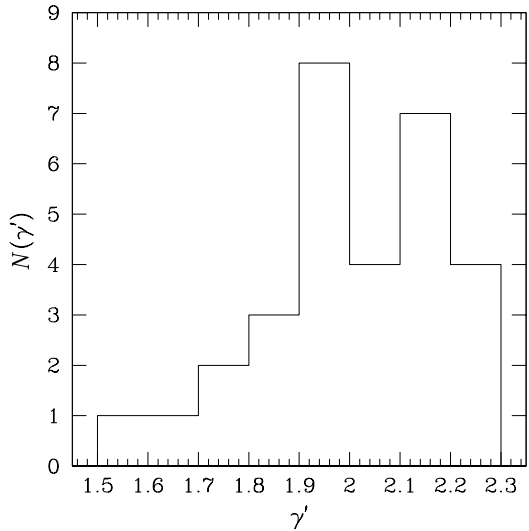}
\caption{Distribution of the values of the best-fit power-law index of
  the total density distribution $\rhotot\propto r^{-\gammap}$ for the
  end-products of all steps of minor and major merging
  hierarchies. The distribution has mean $\langle\gammap\rangle\simeq
  2.01$ and standard deviation $\sigma_{\gammap}\simeq 0.18$.}
\label{fig:histo}
\end{figure}

\begin{figure}
\epsscale{1.0}
\plotone{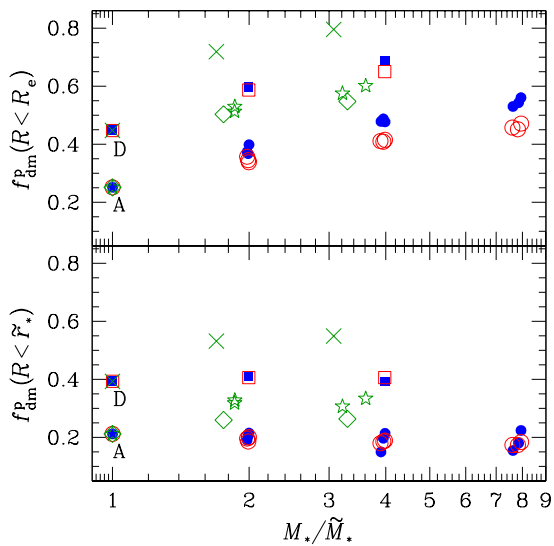}
\caption{Upper panel: projected dark-matter-to-total-matter mass
  ratio within $\Re$ as a function of the stellar mass for the major
  (circles and squares) and minor (stars, diamonds and crosses)
  merging hierarchies. The points representing the seed
  ($\Mstar/\Mstartilde=1$) galaxy models A and D are labelled. For the
  merger remnants $\fdmp$ depends on the line of sight, but projection
  effects are small, so just the average value is plotted.  Lower
  panel: same as upper panel, but plotting projected
  dark-matter-to-total-matter mass ratio within an aperture of radius
  $R=\rstartilde$ (same radius, in physical units, for all models).}
\label{fig:fdm}
\end{figure}


\begin{figure*}
\epsscale{1.}
\plotone{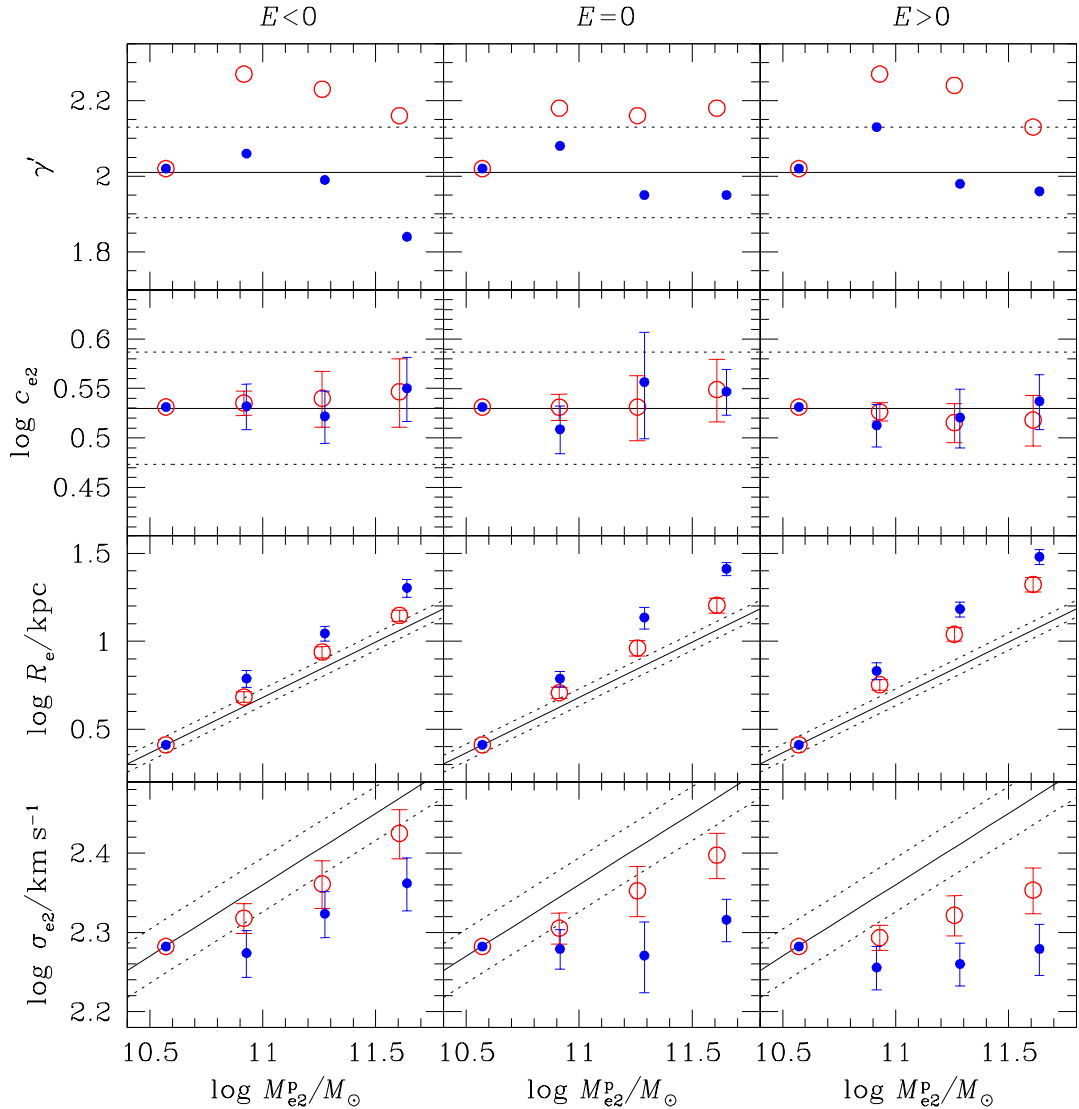}
\caption{From top to bottom: best-fitting logarithmic slope $\gammap$
  ($\rho_{\rm tot}\propto r^{-\gammap}$) of the total density profile
  (first row), structure parameter $\cet$ (second row), effective
  radius (third row) and projected velocity dispersion $\sget$ (fourth
  row), as functions of the projected mass $\Metp$, for elliptic (left
  column), parabolic (central column) and hyperbolic (right column)
  major-merging hierarchies with seed galaxy model A. Points and error
  bars represent the average value and the 1-$\sigma$ scatter (due to
  projection effects) for simulated galaxies: filled circles are for
  head-on encounters and empty circles for off-axis
  encounters. $\Metp$ depends on the line-of-sight, but error bars are
  smaller than the symbol size, so they are not plotted. Solid and
  dotted lines represent the observed correlations with 1-$\sigma$
  scatter.}
\label{fig:all}
\end{figure*}
\begin{figure}
\centerline{ 
\psfig{figure=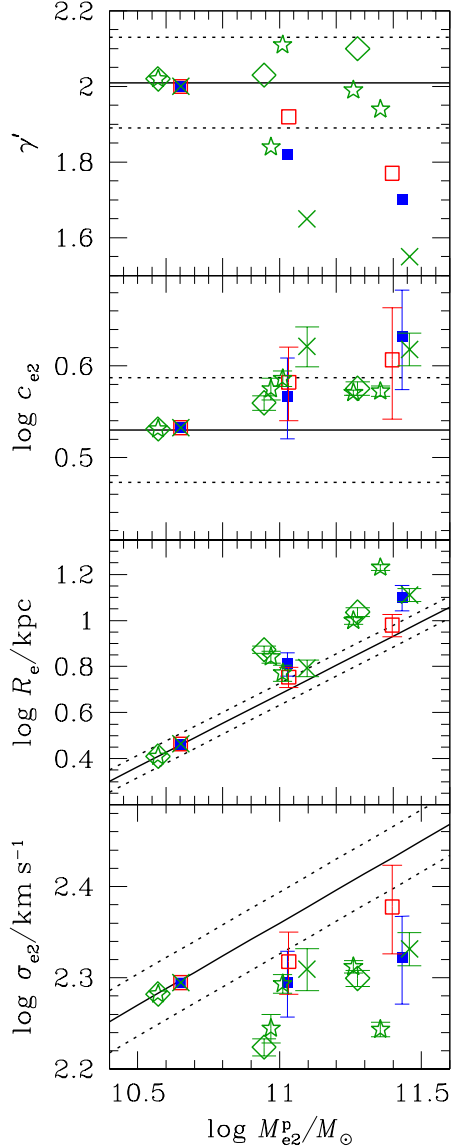,width=0.8\hsize,angle=0,bbllx=40bp,bblly=695bp,bburx=280bp,bbury=150bp,clip=}} 
\caption{Same as Fig.~\ref{fig:all}, but for parabolic major-merger
  hierarchies having as seed galaxy model D (solid squares are for
  head-on encounters and empty squares for off-axis encounters), and
  for the following minor merging hierarchies: model-A hierarchies
  with ten satellites per step (stars) and five satellites per step
  (diamonds), and the model-D hierarchy with ten satellites per step
  (crosses).}
\label{fig:fstar}
\end{figure}

\begin{figure}
\centerline{ 
\psfig{figure=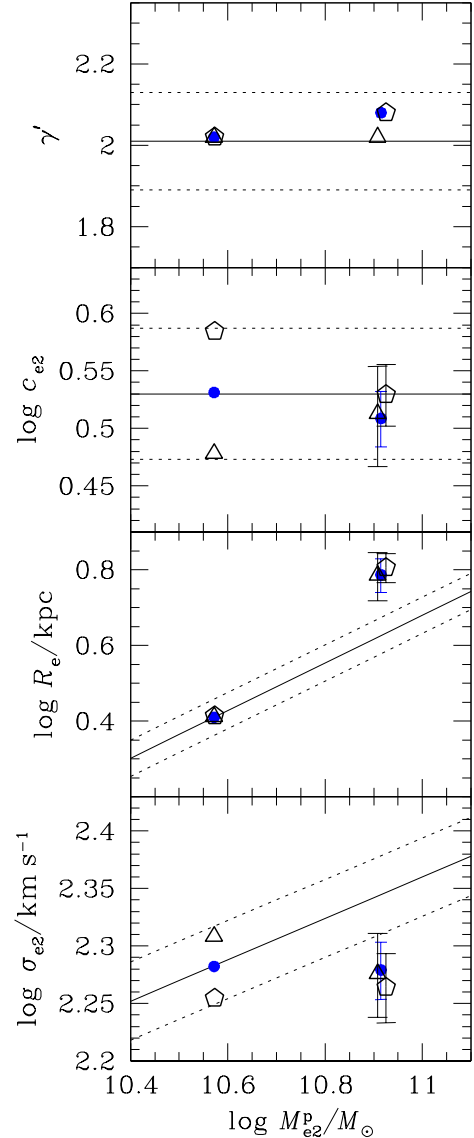,width=0.8\hsize,angle=0,bbllx=40bp,bblly=695bp,bburx=280bp,bbury=150bp,clip=}} 
\caption{Same as Fig.~\ref{fig:all}, but for the seed models and
  end-products of three head-on parabolic major-merging simulations
  differing only for the anisotropy of their seed galaxy models: 2A5ph
  (circles; $\ra/\rstartilde=1.4$), 2B5ph (triangles;
  $\ra/\rstartilde=\infty$) and 2C5ph (pentagons,
  $\ra/\rstartilde=0.7$).}
\label{fig:anis}
\end{figure}

\begin{figure}
\epsscale{1.0}
\plotone{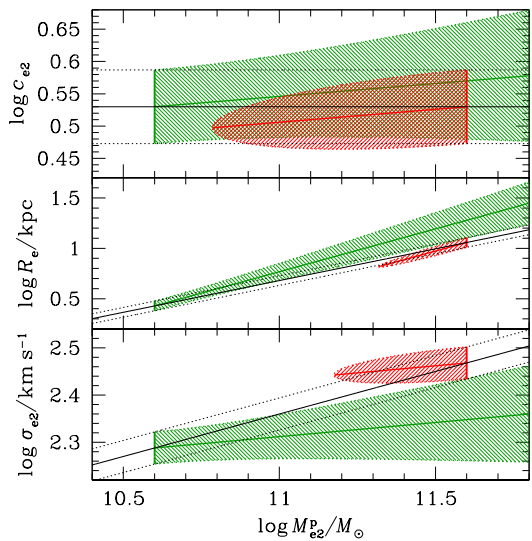}
\caption{ Top: the green shaded area represents the expected
  distribution, within 1 $\sigma$, of the merger remnants of a
  population of galaxies all with the same projected mass
  $\Metp=10^{10.6}\Msun$, and a distribution of $\cet$ matching the
  observed one (see \S~\ref{sec:lens}). The red shaded area represents
  the location in the $\Metp-\cet$ plane of possible dry-merging
  progenitors of a $z=0$ early-type galaxy with projected mass
  $\Metp=10^{11.6}\Msun$ (see \S~\ref{sec:highz}). Central and bottom
  panels: same as top panel, but in the $\Metp-\Re$ and $\Metp-\sget$
  planes. Symbols are the same as in Figs.~\ref{fig:all}
  and~\ref{fig:fstar}. Solid and dotted lines represent the observed
  correlations with 1-$\sigma$ scatter.}
\label{fig:scatt}
\end{figure}

\begin{figure}
\epsscale{1.0}
\plotone{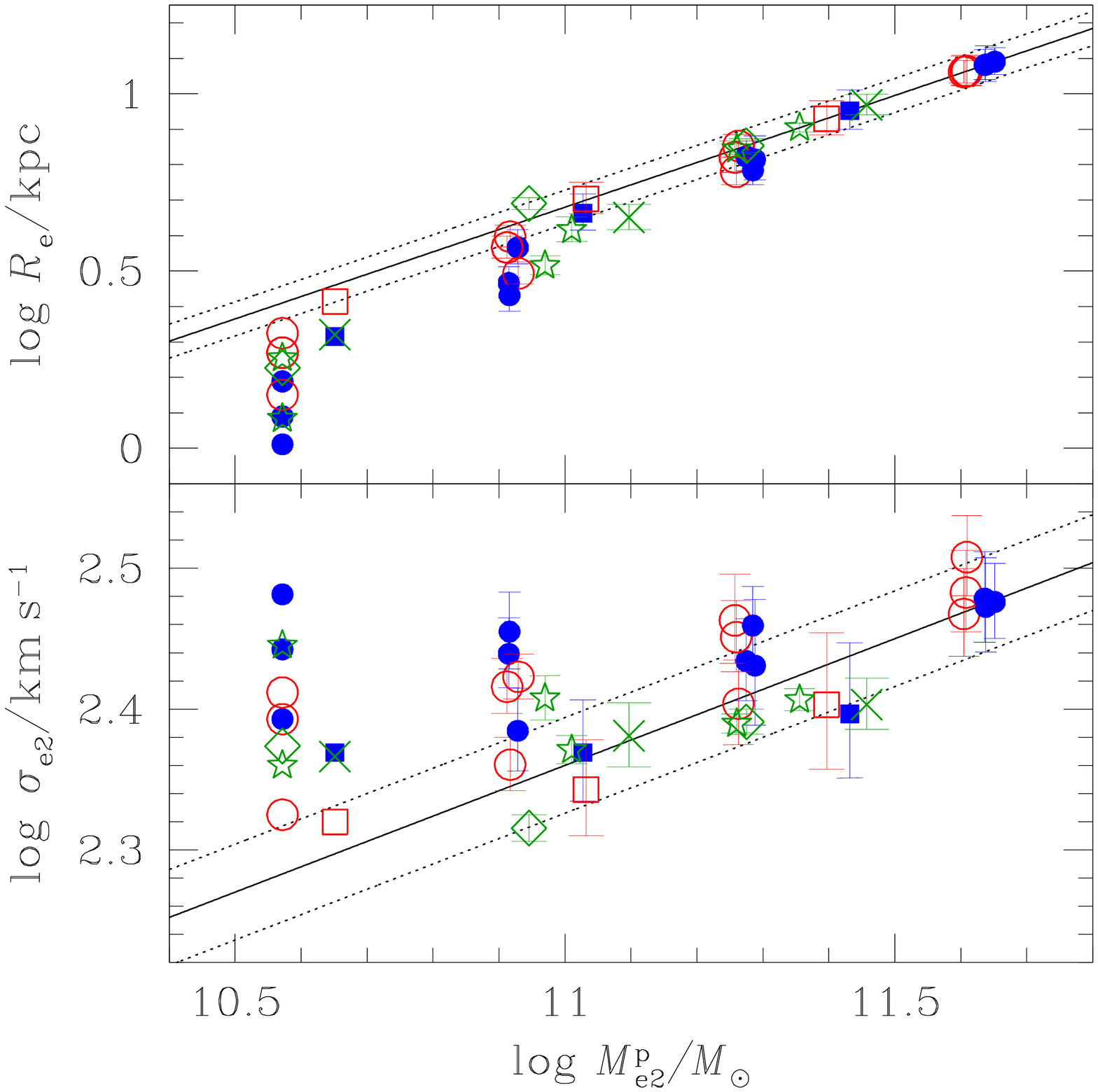}
\caption{Effective radius (upper panel) and projected velocity
  dispersion (lower panel) of the seed galaxies and merger remnants of
  all steps of our major and minor merging hierarchies. For each
  hierarchy, $\Mstartilde$ and $\rstartilde$ are set so that the
  remnant of the last step lies on the local $\Metp-\Re$ relation.
  Solid and dotted black lines represent the local observed
  correlations with 1-$\sigma$ scatter. Symbols are the same as in
  Fig.~\ref{fig:all} and~\ref{fig:fstar}.}
\label{fig:backmp}
\end{figure}

\section{Effect of dry mergers on the mass density distribution}
\label{sec:tden}

\subsection{Total density profiles}

From lensing studies we know that the total density profiles of
early-type galaxies are close to isothermal over a large radial
range. Here we address the question of whether such a feature is
preserved by dry major and minor merging. As a first experiment we
consider the merging between two identical truncated SISs with total
density distribution~(\ref{eqrhosis}). In this simple case each seed
galaxy model is one-component, so we do not distinguish between DM and
baryons. Figure~\ref{fig:sis} plots the total density profiles of the
seed models and of the merger remnant of this run. By construction,
the seed model has density profile $\propto r^{-2}$ within $\sim
\rMtot$ and steeper at larger radii because of the exponential
truncation. The density profile of the merging end-product is very
similar to that of the seed galaxy: fitting it with a power law
$r^{-\gammap}$ in the radial range $0.03<r/\rMtot<1$ we get a
best-fitting index $\gammap=2.00\pm 0.01$. The finding that the {\it
  inner} density cusp is preserved by dissipationless merging is
consistent with theoretical expectations \citep{Deh05} and previous
numerical experiments \citep{Ful01,Kaz06}. Our simulation shows that
{\it the $\sim r^{-2}$ density profile is preserved by dry major
  merging not only in the central regions, but throughout the system.}

Let us now consider how the total density profile is affected by
merging more realistic (two-component) stellar systems.
Figure~\ref{fig:tden} plots, as a function of radius, the
(angle-averaged) total density distributions of our six reference
major merging hierarchies, having as seed galaxy model A and differing
for the orbital parameters of the galaxy encounters.  From the
diagrams it is apparent that the merging process affects significantly
the {\it normalization} of the density profile (producing systems of
lower density at the half-light radius; see also \S~\ref{sec:lens}),
but it roughly preserves its {\it shape}.  To quantify the effect of
dry merging on the shape of the total mass density distribution we fit
the total intrinsic density profiles of our models with a power law
$r^{-\gammap}$ over the radial range $0.1\lsim r/\Re \lsim 1$, where
$\Re$ is the average of the circularized half-light radius over 50
random projections. By construction our seed galaxy models have
close-to-isothermal total density profile: in the radial range
$0.1\lsim r/\Re \lsim 1$ model A and model D are best fitted with
$\gammap=2.02\pm0.02$ and $\gammap=2.00\pm0.01$, respectively.
Figure~\ref{fig:histo} shows the distribution of the best-fitting
$\gammap$ for the end-products of all steps of our merging hierarchies
(30 steps altogether). It is apparent that dry merging can make
slightly steeper or shallower the total density profile, depending on
the characteristics of the galaxy encounters.  The distribution has
mean $\langle\gammap\rangle \simeq 2.01$ and standard deviation
$\sigma_{\gammap}\simeq 0.18$ ($\sim 9\%$). This is comparable with
the observationally determined values $\langle\gammap\rangle \simeq
2.08\pm0.1$ and standard deviation $\sigma_{\gammap}\lesssim0.20$
\citep{Koo09}, suggesting that if a population of early-type galaxies
has initially close-to-isothermal total density profile, dry merging
tends to maintain this property, but adds non-negligible scatter.

The best-fitting values of $\gammap$ for the single simulations are
reported in Table~\ref{tab:sim} together with the other simulation
parameters: off-axis major mergers tend to produce steeper density
profiles than head-on encounters, while there is no evidence of a
dependence of $\gammap$ on the orbital energy. Minor and major merging
involving more DM dominated galaxies (model D) tend to decrease
$\gammap$.

We focused on the slope of the total density profile over the radial
range $0.1\lsim r/\Re \lsim 1$, which is approximately the range
probed in combined strong-lensing and stellar kinematics observations.
However, we note that the total density profiles of our seed models
and merger remnants are well represented by a power-law over a much
larger radial range, though the power-law index is rather sensitive to
the radial interval. For instance, seed models A and D are fitted over
the larger radial range\footnote{Larger radial ranges are interesting
  for comparison with weak-lensing observations of early-type galaxies
  \citep{Gav07}.} $0.1\lsim r/\Re \lsim 10$ with $\gammap=2.17\pm0.02$
and $\gammap=1.94\pm0.02$, respectively. Also fitting over this larger
radial range, the end-products are found to have slightly larger or
smaller best-fitting $\gammap$ than the seed galaxy of their
hierarchy.

\subsection{Central dark-matter fractions}

We have seen that dry merging tends to preserve the slope of the total
density profile. We also found that the shape of the density
distributions of stars and DM is not dramatically affected by dry
mergers (see Fig.~\ref{fig:den}), though there is redistribution
between DM and stars that leads to a systematic increase with merging
of the projected DM-to-total mass ratio $\fdmp$ within $\Re$. This is
apparent from the upper panel Fig.~\ref{fig:fdm}, plotting
$\fdmp(R<\Re)$ as a function of $\Mstar$ for our major and minor
merging hierarchies. The seed galaxy models have $\fdmp=0.25$ (model
A) and $\fdmp=0.45$ (model D), while the merger remnants have DM
fractions in the range $0.3\lsim \fdmp \lsim 0.7$.  In all cases
$\fdmp$ increases with the stellar mass along the merging hierarchy.
Taken by itself this effect might be interpreted as quite naturally
explaining the ``tilt'' seen in the luminosity-based FP relation in
terms of a systematic increase in the central DM fraction with
increasing mass.  If more luminous and massive galaxies represent a
later stage in the merging hierarchy, this increased central DM
fraction will lead to a larger dynamical mass per unit luminosity as
compared to less massive (and less merged) galaxies. However, the
increase of $\fdmp(R<\Re)$ with mass during the merging hierarchy is
mainly driven by a strong increase in the effective radius produced by
dry merging, which might be considered an argument against the
relevance of dry mergers in the assembly of elliptical galaxies (see
\S~\ref{sec:lens}).  In fact, the trend is quite different if one
considers $\fdmp$ as measured within a small aperture of fixed radius,
independent of the galaxy's half-light radius. For instance, in the
lower panel of Fig.~\ref{fig:fdm} we show the results obtained by
measuring the projected DM fraction within and aperture of radius
$R=\rstartilde$. We note that $\rstartilde$ ranges from $\simeq 0.8
\Re$ for the seed galaxies, to $\rstartilde\simeq 0.07 \Re$ for the
most extended remnant.  We find that the remnants have
$\fdmp(R<\rstartilde)$ close the value measured in the seed galaxy of
their hierarchy, indicating that the central projected DM fraction,
measured within a region of fixed size in physical units, is not
substantially affected by dry merging.  Interestingly, minor mergers
(stars, diamonds and crosses in the diagrams) tend to produce systems
with higher values of $\fdmp$ than major mergers (circles and squares
in the diagrams), indicating that dynamical-friction heating
\citep{ElZ04,Nip04,Rom08,Joh09b} is effectively balanced by accretion
of DM in the central regions \citep[see][for a discussion]{Nip04}.


\section{Effect of dry mergers on the lensing scaling relations}
\label{sec:lens}

We address here the question of whether the lensing scaling relations
are preserved by major and minor dry merging. For this reason we
consider seed galaxy models that lie on the observed correlations and
we investigate whether the merger remnants are consistent with the
same relations within their intrinsic scatter. \citet{Nip03a} pointed
out that, when comparing merging simulations with the standard
(luminosity based) scaling relations of early-type galaxies, it is
crucial to consider not only the edge-on FP, but also the
Faber-Jackson and the Kormendy relation, because a system may well lie
on the FP while deviating significantly from the other
relations. Similarly, the fact that a galaxy model lies on the MP does
not mean that it satisfies also the the observed lensing mass-size and
mass-velocity dispersion relations, because ---for fixed $\Metp$---
anomalous values of both $\sget$ and $\Re$ can conspire to give a
value of $\cet$ perfectly consistent with observations. For this
reason we will study how our merger hierarchies behave with respect to
all three lensing scaling laws.

\subsection{Placing the seed galaxies on the observed scaling relations}

As discussed extensively in \citet{Nip08}, for a spherical galaxy
model the dimensionless quantity $\cet$ is fully determined by the
stellar and DM density profiles, and by the distribution of the
orbital velocity of stars. Of course, purely dynamical systems such as
our seed galaxy models can be rescaled arbitrarily by choosing their
mass scale $\Mstartilde$ and length scale $\rstartilde$ in physical
units.  While the value of $\cet$ is independent of this choice, in
order to place a model on the observed mass-size and mass-velocity
dispersion relations, mass and length scales must be chosen
consistently.

Given that the MP relation has no tilt and that the observed mass-size
and mass-velocity dispersion relations are consistent with power laws
over the range covered by the lensing data (equations~[\ref{eq:MP}],
[\ref{eq:reff}] and [\ref{eq:sigma}]), we would not need to specify
the mass of the seed galaxy in order to address the question of the
effect of dry merging. It is sufficient to check whether the merging
end-products have values $\cet$ consistent with the MP and the
relative change of $\Metp$, $\Re$ and $\sget$ are consistent with the
observed mass-size and mass-velocity dispersion relations.  However,
just for clarity and without loss of generality, we fix
$\Mstartilde=10^{11}\Msun$, so all our seed galaxies have total
stellar mass $\Mstar=10^{11}\Msun$.  The projected mass
$\Metp=3.73\times 10^{10}\Msun$ for galaxy model A and
$\Metp=4.48\times 10^{10}\Msun$ for galaxy model D.  We then fix
$\rstartilde=2.08$ kpc for model A and $\rstartilde=2.34$ kpc for
model D, so that they lie on the $\Metp-\Re$ relation (and, as a
consequence, on the $\Metp-\sget$ relation, because they lie on the MP
by construction).  The satellites in the minor-merger hierarchies have
masses and characteristic radii proportionally smaller than those of
the central galaxy (see \S~\ref{sec:min}), but such that they also
satisfy the scaling laws.  Clearly, $\Mstartilde$ and $\rstartilde$
are kept fixed throughout each hierarchy, so masses, sizes and
velocities of all remnants are now known in physical units.

\subsection{Comparing the merger remnants with the observed scaling relations}
\label{sec:remnant}

The location of the merger remnants with respect to the observed
scaling laws is shown in Figs.~\ref{fig:all}, \ref{fig:fstar} and
\ref{fig:anis}. We note that the total projected mass $\Metp$ within
$\Re/2$ increases with dry merging more than the stellar mass or the
total mass integrated over the entire system.  This can be seen, for
instance, in Fig.~\ref{fig:all}: in three merging steps $\Metp$
increases by more than a factor of ten, while the stellar or total
mass, integrated over the entire system, can increase at most by a
factor of eight (in fact less, as a consequence of mass loss). This
finding indicates that dry merging redistributes matter so that the
fraction of mass found in a cylinder of radius $\Re/2$ increases
significantly.

Let us focus now on the behavior of the merger hierarchies with
respect to the MP, which is illustrated in the second row of panels of
Fig.~\ref{fig:all}, \ref{fig:fstar} and \ref{fig:anis}. In general,
the intermediate and final remnants of all our merger hierarchies are
found to be very close to the MP. Actually, all the remnants of
model-A merging hierarchies have values of $\cet$ within the observed
scatter of the MP, while the end-products of the model-D merger
hierarchies (squares and crosses in Fig.~\ref{fig:fstar}) have
slightly higher values of $\cet$, and are only marginally consistent
with the MP. This result implies that $\cet$ is a robust parameter,
since its value is essentially preserved under a wide range of
dry-merging histories.  In a scenario in which early-type galaxies
grow significantly by dry merging, the characteristic value of $\cet$
must therefore be established by the dissipative (``wet'') formation
processes of the primordial progenitor galaxies.  The implications of
the observed value of $\cet$ for galaxy {\em structure} (as opposed to
evolution) are the subject of \citet{Nip08}. Remarkably, dry merging
appears also to wash out the effect of orbital anisotropy on $\cet$:
the end-products of mergers whose seed galaxies differ only in the
velocity distribution of their stars \citep[and thus have different
  values of $\cet$; see][]{Nip08} are indistinguishable in the space
of the parameters $\Metp$, $\Re$ and\, $\sget$ (see
Fig.~\ref{fig:anis}).

As anticipated above, the fact the MP is ``closed'' with respect to
dry merging does not mean that the same must be true also for the
mass--velocity-dispersion and mass--size relations. In fact, as might
be expected from previous explorations of the effect of dry merging on
the Faber-Jackson and Kormendy relations \citep{Nip03a,Boy06}, the
end-products of all the merger hierarchies are found to deviate
significantly from the $\Metp$-$\Re$ and $\Metp$-$\sget$ relations,
because they have too large $\Re$ and too low $\sget$ for their mass
(see third and fourth rows of panels in Fig.~\ref{fig:all},
\ref{fig:fstar} and \ref{fig:anis}). It turns out that the too large
$\Re$ and too low $\sget$ compensate, giving values of $\cet$ very
close to the observed one (we recall that $\Re$ and $\sget$ appear in
the definition of $\cet$ just in the product $\Re\sgetsq$; see
equation~[\ref{eq:cet}]).  Some of the minor-merger hierarchies lead
to strong deviations from the observed scaling relations (see
right-most stars in Fig.~\ref{fig:fstar}): in some steps $\Re$
increases with $\Metp$ more than linearly, and $\sget$ decreases with
$\Metp$, consistent with recent results by \citet{Naab09}. In other
cases the increase in size is significantly smaller, suggesting that
the structural effect of minor merging is quite dependent on the
details of the interaction, such as DM distribution and orbits of the
satellites. In the merging hierarchies having as seed galaxy model D,
whose end-products are found slightly off the MP, the deviations from
the mass-size and mass-velocity dispersion relation are in general
relatively small (squares and crosses in Fig.~\ref{fig:fstar}), and
marginally consistent with the observed relations in the case of
off-axis major mergers (open squares in Fig.~\ref{fig:fstar}). In
these cases low values of $\sget$ are not compensated by large enough
$\Re$.

In the top panels of Fig.~\ref{fig:all}, \ref{fig:fstar} and
\ref{fig:anis} the total-density slope $\gammap$ (see
\S~\ref{sec:tden}) is plotted as a function of mass for the seed
galaxies and the remnants. The behavior on the $\Metp$-$\gammap$ plane
is not strictly correlated to that in the $\Metp$-$\cet$,
$\Metp$-$\Re$ and $\Metp$-$\sget$ planes, though we note that models
that deviate more from the $\Metp$-$\Re$ and $\Metp$-$\sget$
correlations tend to have steeper total density profiles (i.e. higher
values of $\gammap$), while systems with higher values of $\cet$ have
shallower total density profiles.

\subsection{Mass dependence of $\Re$, $\sget$ and $\cet$ for merger remnants}
\label{sec:powlaw}

The growth of galaxies by dry merging of smaller systems can be
quantified by approximating the dependence of the structural
parameters on mass with power-law functions: $\Re\propto
{(\Metp)}^{\alphar}$, $\sget\propto{(\Metp)}^{\alphas}$ and
$\cet\propto{(\Metp)}^{\alphac}$.  For each step of a hierarchy the
power-law indices are given by $\alphar\equiv \Delta \log \Re/\Delta
\log \Metp$, $\alphas\equiv \Delta \log \sget/\Delta \log \Metp$ and
$\alphac\equiv \Delta \log \cet/\Delta \log \Metp$, where $\Delta$
indicates variation between the remnant and the progenitor
[e.g. $\Delta \log \Re = (\log\Re)_{\rm remnant}-(\log\Re)_{\rm
    progenitor}$].  We can obtain an estimate of the expected global
effect of a general dry-merging hierarchy by considering the average
values ($\langle\alphar\rangle=0.85$, $\langle\alphas\rangle=0.06$,
$\langle\alphac\rangle=0.04$) and the associated standard deviations
($\delta\alphar=0.17$, $\delta\alphas=0.08$, $\delta\alphac=0.07$) of
these values for our major and minor merging hierarchies (30 steps
altogether).  These values must be compared with the slope of the
observed scaling laws (equations~[\ref{eqobsrange}], [\ref{eq:reff}]
and [\ref{eq:sigma}]), which correspond to $\alphar=0.63$,
$\alphas=0.18$ and $\alphac=0$, consistent with our statement that the
MP is roughly preserved by dry mergers, but the mass--size and
mass--velocity-dispersion relations are not.

Remarkably, the distributions of $\alphar$, $\alphas$ and $\alphac$
for our N-body experiments are characterized by a substantial scatter,
which is quantified by the quoted values of the standard
deviations. Such a broad distribution around the average slope is an
additional problem for the dry-merging scenario, because it implies
that dry mergers {bf would introduce a lot of dispersion in the
  scaling relations}. To quantify this, let us consider a population
of galaxies, all with $\Metp=M_0$ and ${\Re}=R_0$, that evolve through
dry merging with an unspecified merging history made of both minor and
major mergers (we are not assuming that these seed galaxies merge just
among themselves): under the hypothesis that $\alphar$ is distributed
normally, the merger remnants of mass $\Metp>M_0$ are expected to have
$\log\Re$ normally distributed with average
$\langle\log\Re\rangle=\log R_0 +
\langle\alphar\rangle\log(\Metp/M_0)$ and standard deviation
$\delta\alphar\log(\Metp/M_0)$. Thus, in this simple example, the
predicted scatter in the $\Metp-\Re$ relation at a given $\Metp$
increases for increasing $\delta\alphar$, and ---for fixed
$\delta\alphar$--- increases for increasing $\Metp$, suggesting that a
large enough mass growth by dry merging would produce a spread in
$\Re$ at given $\Metp$ inconsistent with the tight observed
correlation.  The importance of this effect can be appreciated by
looking at the green shaded area in the central panel of
Fig.~\ref{fig:scatt}, corresponding to the expected distribution,
within 1 $\sigma$, of the merger remnants of a population of galaxies
all with the same projected mass $\Metp=10^{10.6}\Msun$ (green
vertical bar in the diagram), but now with a more realistic
distribution of $\Re$: we assume that $\log \Re$ is normally
distributed with average $\langle\log\Re\rangle$ satisfying the
observed $\Metp$-$\Re$ correlation~(equation~[\ref{eq:reff}]) and
standard deviation equal to its intrinsic scatter ($0.048$ dex).
Similar considerations apply to $\cet$ and $\sget$ (green shaded areas
in the top and bottom panels of Fig.~\ref{fig:scatt}). It is apparent
that the scatter in $\Re$ and $\sget$ for the remnant population
rapidly increases with mass above the observed scatter, while the
effect is weaker in the $\Metp$-$\cet$ plane.  The exact shapes of the
green shaded areas depend on the set of merging hierarchies
considered. For instance, in the $\Metp$-$\Re$, we find a weaker
deviation from the observed correlation if we consider only model-D
hierarchies, and a stronger deviation if we consider only
minor-merging hierarchies (as a consequence of the specific behavior
of these models; see \S~\ref{sec:remnant}): in any case regions of the
diagrams out of the observed strip are quickly populated.

Summarizing, we showed that the lensing mass--size and
mass--velocity-dispersion relations are effectively destroyed by dry
mergers, which produce systems with too large half-light radii and too
low velocity dispersions, and introduce substantial scatter in $\Re$
and $\sget$ at given mass. 
We conclude that {\it present-day
  early-type galaxies did not form by dry merging of high-$z$
  early-type galaxies obeying the local ($z=0$) lensing scaling
  relations}.

\section{Dry mergers and redshift evolution of early-type galaxies}
\label{sec:highz}

In the previous section we explored the effect of dry mergers on
galaxies lying on the lensing scaling relations observed locally.
However, higher-$z$ early-type galaxies must not necessarily obey the
$z=0$ lensing scaling laws. In fact, though the lensing scaling
relations at higher redshift are not available, measures of the
effective radii and the stellar masses of high-redshift ($z\gsim 1-2$)
early-type galaxies suggest that these objects may be remarkably more
compact than their local counterparts
\citep[e.g.,][]{Dad05,Tru06,Zir07,Cimatti08,vDo08,vdW08}.  As shown by
our current results, as well as previous work
\citep[][]{Nip03a,Boy06}, dry merging has the effect of making
galaxies less compact, so it has been proposed that evolution of
early-type galaxies through such a mechanism might be consistent with
the recent observational finding that higher-$z$ galaxies are more
compact \citep[e.g.,][]{Kho06,Hop09,vdW09}.  Even allowing for the
possibility that the candidate dry-merging progenitors of present-day
early-type galaxies do not obey the local lensing scaling laws, the
tightness of these correlations can be used to constrain the
contribution of dry mergers in the assembly history of early-type
galaxies.

For this purpose it is useful to consider the following question:
under the hypothesis that local early-type galaxies (satisfying the
lensing scaling laws) formed by dry merging, what are the expected
properties of their (high-redshift) progenitors?  To address this
question we assume that the remnants of the last steps of all our
simulated merging hierarchies satisfy the local $\Metp-\Re$
relation. We then fix $\Mstartilde=10^{11}\Msun$, and we choose a
different value of $\rstartilde$ for each hierarchy (in all cases
smaller than the values of $\rstartilde$ adopted
in~\S~\ref{sec:lens}), so $\Metp$, $\Re$ and $\sget$ are given in
physical units for all models.  The location of the models with
respect to the observed lensing correlations is shown in
Fig.~\ref{fig:backmp}: the most massive systems lie (by construction)
on the observed $\Metp-\Re$ relation (upper panel) and approximately
also on the $\Metp-\sget$ relation (lower panel), consistent with the
fact that $\cet$ is not significantly affected by dry merging; less
massive systems lie systematically below the $\Metp-\Re$ relation and
above the $\Metp-\sget$ relation.  In this picture the lowest-mass
systems in the diagrams in Fig.~\ref{fig:backmp} (those with
$\log\Metp/\Msun\lsim 10.7$, all having by construction
$\Mstar=10^{11}\Msun$) would represent the high-$z$ progenitors of
local ellipticals. Given the idealized nature of our experiments,
which are not set in a full cosmological context, we cannot define
precisely the redshift of these progenitors. However, a lower limit to
their redshift can be obtained by noting that the time to complete our
merging hierarchies is in the range $6-12$ Gyr, with median $\sim 9$
Gyr. A realistic merging hierarchy is expected to be a mixture of
steps of the explored hierarchies, so we take as reference time the
median, which is the look-back time to $z\sim 1.4$. Therefore, in the
present context, our seed galaxy models should be compared with
$z\gsim1.4$ early-type galaxies with stellar mass $\sim
10^{11}\Msun$. The diagrams in Fig.~\ref{fig:backmp} show that the
candidate progenitors are predicted to be compact, with effective
radii up to a factor of $\sim 2.5$ smaller and velocity dispersions up
to a factor of $\sim 1.6$ higher than local galaxies with similar
mass, and are characterized by a significant spread in both $\Re$ and
$\sget$. We recall that the candidate progenitors have values of
$\cet$ consistent with those observed in present-day galaxies, because
our seed galaxies lie on the MP by construction.  Though one cannot
assume a priori that high-z progenitors lie on the MP, this assumption
is justified as far as we are assuming ---as working hypothesis---
that local early-type galaxies formed by dry merging, because we have
shown that dry mergers tend to preserve the MP.

The average properties of the candidate progenitors are qualitatively
consistent with those of $z\gsim 1.4$ early-type galaxies
\citep[see][]{Cimatti08,Cap09}, so one might be tempted to conclude
that a dry-merging scenario is compatible with observations. This is
not necessarily the case, because the tightness of the local relations
must be reproduced as well. In fact, as we are going to show, the
tightness of the local relations sets the strongest constraints on dry
merging, requiring a remarkable degree of fine tuning of the mix of
progenitors and types of interaction.

This can be seen from the example shown in Fig.~\ref{fig:scatt}, where
we focus, without loss of generality, on local early-type galaxies
with projected mass $\Metp=10^{11.6}\Msun$ (red vertical bars
in the diagrams), and we constrain the properties of their progenitors
in a dry-merging scenario.  We have seen in \S~\ref{sec:powlaw} that
progenitors of given $\Metp$, $\sget$ and $\Re$ would produce through
dry merging a population of galaxies with a range of values of
$\sget$, $\Re$ and $\cet$ for given mass, depending on the expected
distributions of $\alphar$, $\alphas$ and $\alphac$, which are
constrained by our N-body simulations. Here we take the opposite
approach and impose that our reference present-day galaxies with
$\Metp=10^{11.6}\Msun$ have $\log\cet$, $\log\Re$ and
$\log\sget$ with average values and intrinsic scatter consistent with
observations.  This procedure delivers constraints on $\Metp$,
$\sget$, $\Re$ and $\cet$ in the form of ``allowed'' regions in the
diagrams of Fig.~\ref{fig:scatt} (red shaded areas). For instance, let
us focus on the $\Metp-\Re$ plane (central panel): at a given mass
$\Metp<10^{11.6}\Msun$, the upper and lower limits of the red
shaded areas bracket the distribution within $1\sigma$ of $\log\Re$ of
a normally distributed population of galaxies of mass $\Metp$ that can
produce via dry merging our population of galaxies with lensing mass
$10^{11.6}\Msun$.  The same considerations apply to the
$\Metp$-$\cet$ and $\Metp$-$\sget$ planes.  It is clear that the
smaller the shaded area, the higher the fine tuning necessary to
obtain the local distributions.

While the constraints in the $\cet$ plane are weak (because $\cet$ is
roughly preserved by dry merging), those in the $\sget$ and $\Re$
plane are quite strong and indicate that an uncomfortable degree of
fine tuning is required in a dry merging scenario. Note, in
particular, that a growth in $\Metp$ via dry merging by more than a
factor of $\sim 2$ (and in size by more than a factor of $\sim 1.8)$
is excluded (see the red shaded area in the $\Metp$-$\Re$
plane). Though our results constrain the projected mass $\Metp$
measured within $\Re/2$, the upper limits on the mass growth can be
extended to the total mass, because in our simulations $\Metp$
increases with dry merging faster than the total mass (integrated over
the entire system; see \S~\ref{sec:remnant}). Therefore, our results
indicate that present-day, massive early-type galaxies did not
assembled more than $50\%$ of their total (dark plus luminous) mass
through dry merging.

One important caveat is that the allowed amount of mass growth depends
on $\langle\alphar\rangle$ and $\delta\alphar$, and thus on the
considered merging histories. Our set of simulations is biased in
favour of major mergers and our exploration of the orbital parameter
space is limited, especially in the case of the more DM dominated
(model D, $\fstar=0.02$) galaxies. For instance, if we consider only
our minor-merger hierarchies the upper limit to the mass fraction goes
down to $\sim 40\%$, while this figure is $\sim 60 \%$ if we consider
only model-D hierarchies. In any case, a certain amount of fine tuning
is required.  Larger sets of numerical experiments, based on more
realistic merging histories, are needed to quantify whether the amount
of fine tuning required is consistent with hierarchical growth in a
cosmological context.

From Fig.~\ref{fig:scatt} it is also apparent that even an increase in
mass by a factor smaller than 2 requires a lot of fine tuning, in the
sense that the progenitors must have tighter mass--size and
mass--velocity-dispersion correlations than local galaxies. This
cannot be excluded, because there is no information so far on the
lensing scaling relations beyond the redshift range covered by the
SLACS sample ($z\sim0.2-0.3$).

\section{Conclusions}
\label{seccon}

We explored the effect of minor and major dry mergers on the lensing
scaling relations of early-type galaxies.  Our main findings can be
summarized as follows.  Major and minor dry mergers: 

\begin{enumerate}
\item Preserve the nearly isothermal structure ($\rho_{\rm tot}\propto
  r^{-2}$) of early-type galaxies, adding a scatter of $\sim 9\%$ to
  the logarithmic slope (for an increase in stellar mass of up to a
  factor of $\sim 8$), consistent with the observed value of
  $\sim$10\%.

\item Do not change the ratio between total (lensing) mass $\Metp$ and
  ``virial'' mass ($\Re\sgetsq/2G$) more than the observed scatter of
  0.057 dex. In other words, they move galaxies along the MP.

\item Move galaxies away from the observed correlations between total
  (lensing) mass and size or velocity dispersion. Specifically, dry
  mergers increase the radius more rapidly than the scaling relation
  with mass would predict, while they do not increase velocity
  dispersion rapidly enough.

\item Add substantial scatter in the lensing mass-size and
  mass--velocity-dispersion relations.

\end{enumerate}

The first two findings indicate that two important properties of
early-type galaxies ---the isothermal total mass profile and the MP---
are quite robust against dry merging.  If these regular structural and
kinematic properties are established ---presumably at $z>1$, in order
to satisfy constraints on the old age of their stellar populations
\citep[e.g.,][]{Tre05} and of the lack of evolution of the logarithmic
slope of the total mass density profile \citep{Koo06,Koo09}---
further growth via dry merging would not spoil this structural and
kinematic homology of early-type galaxies.

The third and fourth findings pose a significant problem for a
scenario where early-type galaxies grow significantly in mass via dry
mergers.  The non-conservation of the mass--size and
mass--velocity-dispersion relation shows that the progenitors of
present day early-type galaxies cannot be obeying the same scaling
relations. On average, they will have to be smaller in size for a
given velocity dispersion or total mass, qualitatively in line with
some recent observational results \citep{vdW08}. However, and most
importantly, dry mergers add substantial scatter to the scaling
relations. A few random generations of dry mergers would be sufficient
to produce a scatter much larger than observed.  In particular, our
results exclude that present-day massive early-type galaxies assembled
more than $\sim 50\%$ of their total (luminous plus dark) mass and
increased their size by more than a factor $\sim 1.8$ via dry merging.
Quantitatively, the amount of dispersion introduced in the scaling
laws by dry merging depends on the detailed properties of the merging
history, such as DM content of the progenitors and mass-ratio and
orbital parameters of the galaxy encounters.  A high degree of fine
tuning - where the properties of the progenitor correlate with the
peculiarities of the mergers process -- would be needed to reconcile
dry mergers with the tight observed lensing scaling relations.

In the present work we considered only gas-free mergers, so our
results do not apply if mergers occur in the presence of significant
amounts of gas. It is well known that dissipative processes might help
reconcile merging with the standard scaling relations
\citep[e.g.][]{Robertson06,Ciotti07,Joh09a}, so it is reasonable that
the same can happen for the lensing scaling relations. However, the
role of wet (gas-rich) mergers in the formation of early-type
galaxies, though not strongly constrained by the scaling relations, is
limited by other considerations: if these systems experienced a
significant amount of mergers in relatively recent times, most of
these mergers must have been dry, because the old stellar populations
of early-type galaxies \citep[e.g.][]{Thomas05} are inconsistent with
substantial recent star formation. It is also the case that in very
massive galaxies most accreted cold gas is likely to be eliminated by
evaporation from the hot interstellar medium before it can form stars
\citep{Nip07}.

We emphasize that our constraints on the dry-merging history apply to
massive early-type galaxies, of which the SLACS sample is
representative, and are not in conflict with the hypothesis that
exceptionally luminous, brightest cluster galaxies (BCGs) formed via
multiple dissipationless mergers, the so-called galactic cannibalism
scenario \citep{Ost75,Hau78}. In fact, numerical simulations of
dynamical-friction driven galactic cannibalism predict that brightest
cluster galaxies must be more extended than expected from the
extrapolation of the size-luminosity correlation of smaller galaxies
\citep{Nip03b,Rus09}, as observed \citep[e.g.,][]{Oeg91,Ber09}. In
this respect, it will be very interesting to perform a systematic
exploration of the behavior of brightest cluster galaxies with respect
to the lensing scaling relations explored here. Larger numbers of BCGs
with lensing features and stellar kinematics than those currently
available \citep[e.g.,][]{S08} are needed to carry out this study.

In conclusion, we can see three main ways to make progress and
determine whether the tightness of the lensing scaling relations is
fatal to the idea of substantial growth of normal (i.e. non-BCG)
early-type galaxies by dry mergers. From a theoretical point of view,
fully cosmological simulations are needed to correlate properties of
the progenitors with the details of the merging process and verify
whether the required extraordinary amount of fine tuning does indeed
occur. Also, future studies need to incorporate stellar mass as a
fundamental variable that can also be directly linked to
observations. From an observational point of view, large and
homogeneous samples of lenses at higher-$z$ than SLACS are needed to
determine how the lensing scaling relations evolve with redshift.

\acknowledgments

We are grateful to our friends and collaborators on the SLACS project
(M.W.~Auger, S.C.~Burles, M.~Barnab\'e, O.~Czoske, R.~Gavazzi,
L.V.E.~Koopmans, P.J.~Marshall, L.A.~Moustakas, S.~Vegetti) for many
stimulating scientific conversations, and for their fundamental role
in putting together the dataset that inspired this work. C.N. is happy
to thank B.~Nipoti for helpful discussions. Support for the SLACS
project (HST-GO programs \#10174, \#10587, \#10886, \#10494, \#10798,
\#11202) was provided by NASA through a grant from the Space Telescope
Science Institute, which is operated by the Association of
Universities for Research in Astronomy, Inc., under NASA contract NAS
5-26555.  T.T.  acknowledges support from the NSF through CAREER award
NSF-0642621, by the Sloan Foundation through a Sloan Research
Fellowship, and by the Packard Foundation through a Packard
Fellowship.  Some of the numerical simulations were performed using
the BCX system at CINECA, Bologna, with CPU time assigned under the
INAF-CINECA agreement 2008-2010.



\begin{deluxetable}{llcccccccc}
\tablewidth{0pt}
\tablecaption{Seed galaxy models.}
\tablehead{
\colhead{Id}            & \colhead{Model}      &
\colhead{$\ra/\rstartilde$}          & \colhead{$\Conc$}  &
\colhead{$\xi$}          & \colhead{$\Mdm/\Mstartilde$}    & \colhead{$\Ntot/10^5$} & $\gammap$ }
\startdata 
A5   & A & 1.4       & 7 & 7.8   & 10&  $11$ & 2.02  \\
A1   & A & 1.4       & 7 & 7.8   & 10 & $2.2$& 2.02 \\
B5   & B & $\infty$  & 7 & 7.8   & 10 & $11$ & 2.02 \\
C5   & C & 0.7       & 7 & 7.8   & 10 &  $11$& 2.02 \\
D1   & D & 1.0       & 8 & 11.6   & 49 & $10$ & 2.00 \\
\enddata
\tablecomments{Id: name of the N-body realization. Model: name of the
  seed galaxy model. $\ra$: anisotropy radius. $\Conc$: NFW
  concentration. $\xi=\rs/\Re$. $\Mdm$: total dark-matter
  mass. $\Mstartilde$: total stellar mass. $\Ntot$: total number of
  particles in the N-body model (stellar and dark matter particles
  have the same mass).  $\gammap$: logarithmic slope of the best-fit
  power-law total density profile ($\rhotot\propto r^{-\gammap}$) over
  the radial range $0.1\leq r/\Re \leq 1$. }
\label{tab:mod}
\end{deluxetable}

\begin{deluxetable}{llccccccccccc}
\tablewidth{0pt}
\tablecaption{Parameters of the simulations and properties of the end-products.}
\tablehead{
\colhead{Id} & \colhead{Prog.}                &
\colhead{$\hatE$}    &  \colhead{$\hatL$} &
\colhead{$\dz$}    &  \colhead{$\vzeropar$} & \colhead{$\vzeroperp$} & \colhead{$\Mstarfin$}
 & \colhead{$\Mdmfin$} &  \colhead{$\sigmav$} & \colhead{$\rMtot$}  & \colhead{$\gammap$}  }
\startdata   
2A5ph & A5    &  0    & 0    & 140 & 0.560 & 0     & 1.99 & 18.66 & 0.493 & 40.04 & 2.06\\
2A5po & A5    &  0    & 1.78 & 140 & 0.542 & 0.140 & 1.99 & 18.41 & 0.491 & 41.21 & 2.16\\
2B5ph & B5    &  0    & 0    & 140 & 0.560 & 0     & 1.99 & 18.99 & 0.486 & 41.22 & 2.08\\
2C5ph & C5    &  0    & 0    & 140 & 0.560 & 0     & 1.99 & 19.03 & 0.495 & 39.23 & 2.02\\
2A1ph & A1    &  0    & 0    & 140 & 0.560 & 0     & 1.99 & 19.01 & 0.490 & 40.28 & 2.08\\
2A1po & A1    &  0    & 1.78 & 140 & 0.542 & 0.140 & 1.99 & 18.76 & 0.492 & 41.17 & 2.18\\
2A1eh & A1    & -0.83 & 0    & 140 & 0.336 & 0     & 2.00 & 19.73 & 0.525 & 33.77 & 2.06\\
2A1eo & A1    & -0.83 & 1.78 & 140 & 0.305 & 0.140 & 2.00 & 19.62 & 0.535 & 35.65 & 2.27\\
2A1hh & A1    & 0.83  & 0    & 140 & 0.716 & 0     & 1.98 & 17.91 & 0.459 & 42.69 & 2.13\\
2A1ho & A1    & 0.83  & 1.78 & 140 & 0.702 & 0.140 & 1.98 & 17.59 & 0.467 & 40.98 & 2.27\\
2D1ph & D1    & 0     & 0    & 240 & 0.910 & 0     & 1.99 & 92.66 & 0.746 & 72.79 & 1.82\\
2D1po & D1    & 0     & 1.78 & 240 & 0.886 & 0.218 & 1.99 & 92.55 & 0.743 & 75.88 & 1.92\\
4A1ph & 2A1ph &  0    & 0    & 280 & 0.548 & 0     & 3.95 & 36.52 & 0.526 & 69.49 & 1.95\\
4A1po & 2A1po &  0    & 1.78 & 280 & 0.528 & 0.130 & 3.95 & 35.28 & 0.530 & 85.17 & 2.16\\
4A1eh & 2A1eh & -0.83 & 0    & 280 & 0.286 & 0     & 3.99 & 38.72 & 0.613 & 55.72 & 1.99\\
4A1eo & 2A1eo & -0.83 & 1.78 & 280 & 0.238 & 0.120 & 3.99 & 38.38 & 0.583 & 63.14 & 2.23\\
4A1hh & 2A1hh &  0.83 & 0    & 280 & 0.678 & 0     & 3.90 & 32.10 & 0.473 & 90.19 & 1.98\\
4A1ho & 2A1ho &  0.83 & 1.78 & 280 & 0.668 & 0.122 & 3.90 & 29.51 & 0.481 & 87.52 & 2.24\\
4D1ph & 2D1h  &  0    & 0    & 480 & 0.888 & 0     & 3.89 & 176.93 & 0.803 & 134.09 & 1.70\\
4D1po & 2D1o  &  0    & 1.78 & 480 & 0.864 & 0.208 & 3.98 & 175.23 & 0.804 & 152.65 & 1.77\\
8A1ph & 4A1ph &  0    & 0    & 560 & 0.538 & 0     & 7.85 & 70.78 & 0.550 & 160.47 &  1.95\\
8A1po & 4A1po &  0    & 1.78 & 560 & 0.509 & 0.144 & 7.82 & 65.83 & 0.528 & 201.62 &  2.18\\
8A1eh & 4A1eh & -0.83 & 0    & 560 & 0.066 & 0     & 7.95 & 76.15 & 0.679 & 84.12  &  1.84\\
8A1eo & 4A1eo & -0.83 & 1.78 & 560 & 0.086 & 0.114 & 7.95 & 75.46 & 0.660 & 106.01 &  2.16\\
8A1hh & 4A1hh &  0.83 & 0    & 560 & 0.666 & 0     & 7.63 & 55.91 & 0.490 & 169.73 &  1.96\\
8A1ho & 4A1ho &  0.76 & 0.80 & 560 & 0.640 & 0.060 & 7.60 & 50.44 & 0.491 & 184.16 &  2.13\\
2A5m10x & A5    &  -    & -    & -   & -     & -     & 1.86 & 15.81 & 0.587 & 22.64  &  2.11\\
2A1m10x & A1    &  -    & -    & -   & -     & -     & 1.86 & 16.09 & 0.585 & 23.02  &  2.11\\
2A1m10y & A1    &  -    & -    & -   & -     & -     & 1.86 & 15.66 & 0.554 & 25.28  &  1.84\\
2A1m5   & A1    &  -    & -    & -   & -     & -     & 1.76 & 14.67 & 0.550 & 23.87  &  2.03\\
2D1m10  & D1    &  -    & -    & -   & -     & -     & 1.69 & 73.13 & 0.896 & 37.74  &  1.65\\
4A1m10x & 2A1m10x &  -    & -    & -   & -     & -     & 3.21 & 26.99 & 0.642 & 32.01  &  1.99\\
4A1m10y & 2A1m10y &  -    & -    & -   & -     & -     & 3.61 & 30.18 & 0.557 & 48.47  &  1.94\\
4A1m5 & 2A1m5 &  -    & -    & -   & -     & -     & 3.29 & 26.69 & 0.597 & 38.83  &  2.10\\
4D1m10  & 2D1m10    &  -    & -    & -   & -     & -     & 3.07 & 119.29 & 0.993 & 50.07  &  1.66 \\
\enddata
\tablecomments{ Id: name of the run and of the corresponding
  end-product. Prog.: name of the progenitor. $\hatE$: dimensionless
  orbital energy. $\hatL$: dimensionless orbital angular momentum
  modulus. $\dzero$: initial
  separation. $\vzeropar\equiv\vvzero\cdot\dvzero/\dzero$: parallel
  component of the initial relative velocity.
  $\vzeroperp\equiv||\vvzero\times\dvzero||/\dzero$: orthogonal
  component of the initial relative velocity. $\Mstarfin$: final
  stellar mass. $\Mdmfin$: final dark matter mass. $\sigmav$: virial
  velocity dispersion of the end-product. $\rMtot$: circularized
  half-mass radius of the total density distribution of the
  end-product. $\gammap$: logarithmic slope of the best-fit power-law
  total density profile ($\rhotot\propto r^{-\gammap}$) over the
  radial range $0.1\leq r/\Re \leq 1$.  Masses, lengths and velocities
  are in units of $\Mstartilde$, $\rstartilde$ and $\vstartilde$,
  respectively.  The parameters $\hatE$, $\hatL$, $\dz$, $\vzeropar$
  and $\vzeroperp$ are defined only for binary major-merging
  simulations.}
\label{tab:sim}
\end{deluxetable}


\end{document}